\documentclass[english,prl,aps,twocolumn,10pt]{revtex4-1}

\usepackage{amsmath}
\usepackage{amssymb}
\usepackage{graphicx}
\usepackage{babel}
\usepackage{array}
\usepackage{verbatim}
\usepackage[colorlinks=true, pdfstartview=FitV, linkcolor=blue, citecolor=blue, urlcolor=blue]{hyperref} % enable links

\def\C        {{$^{13}$C \/}}

\newcolumntype{L}[1]{>{\raggedright\arraybackslash}p{#1}}
\newcolumntype{C}[1]{>{\centering\arraybackslash}p{#1}}

\newcommand{\captionstyle}{\normalfont} %define a caption font size

\newcommand{\mr}[1]{\mathrm{#1}}
\newcommand{\unit}[1]{\,\mathrm{#1}}

\newcommand{\uT}{\,\mu{\rm T}}

\newcommand{\apar}{a_{\mathbin{\|}}}
\newcommand{\aperp}{a_{\mathbin{\bot}}}
\newcommand{\Bac}{B_\mr{ac}}

\newcommand{\fac}{f_\mr{ac}}
\newcommand{\df}{\delta f}

\newcommand{\ts}{t_s}
\newcommand{\tpi}{t_\pi}
\newcommand{\dt}{\delta t}

\newcommand{\dy}{\delta y}

\begin{document}

\title{High resolution quantum sensing with shaped control pulses}

\author{J. Zopes$^1$, K. Sasaki$^2$, K. S. Cujia$^1$, J. M. Boss$^1$, K. Chang$^1$, T. F. Segawa$^1$, K. M. Itoh$^2$, and C. L. Degen$^1$}
\affiliation{$^1$Department of Physics, ETH Zurich, Otto Stern Weg 1, 8093 Zurich, Switzerland.}
\affiliation{$^2$School of Fundamental Science and Technology, Keio University, Yokohama 223-8522, Japan.}

\email{degenc@ethz.ch}

\begin{abstract}
We investigate the application of amplitude-shaped control pulses for enhancing the time and frequency resolution of multipulse quantum sensing sequences.
Using the electronic spin of a single nitrogen vacancy center in diamond and up to 10,000 coherent microwave pulses with a cosine square envelope, we demonstrate 0.6\,ps timing resolution for the interpulse delay.  This represents a refinement by over 3 orders of magnitude compared to the 2\,ns hardware sampling.  We apply the method for the detection of external AC magnetic fields and nuclear magnetic resonance signals of $^{13}\mathrm{C}$ spins with high spectral resolution.
%Our method is simple to implement and useful for many applications that require precise pulse timing.
Our method is simple to implement and especially useful for quantum applications that require fast phase gates, many control pulses, and high fidelity.
\end{abstract}

\date{\today}

\maketitle

Pulse shaping is a well-established method in many areas of physics including magnetic resonance \cite{bauer84,murdoch87,baum85}, trapped ion physics \cite{choi14,timoney08} and superconducting electronics \cite{kelly14} for improving the coherent response of quantum systems. Introduced to the field of nuclear magnetic resonance (NMR) spectroscopy in the  1980s, shaped pulses enable selective excitation of nuclear spins with uniform response over the desired bandwidth, which led to more selective and more sensitive measurement techniques.  More recently, with the advent of fast arbitrary waveform generators (AWG), pulse shaping techniques also entered the field of electron paramagnetic resonance (EPR), thereby improving spectrometer performance via chirped pulses for broadband excitation \cite{segawa15,doll16}.  In quantum information processing applications, shaped microwave pulses are utilized to optimize the fidelity and stability of gate operations against environmental perturbations that cause detuning of transition frequencies or fluctuations in the driving frequency \cite{chow10,cross15}.
%In quantum information processing applications, for example with superconducting qubits, shaped microwave pulses are utilized to optimize the fidelity and stability of gate operations against environmental perturbations that cause detuning of transition frequencies or fluctuations in the driving frequency \cite{chow10,cross15}.

%
\begin{figure}[t]
\includegraphics[width=0.92\columnwidth]{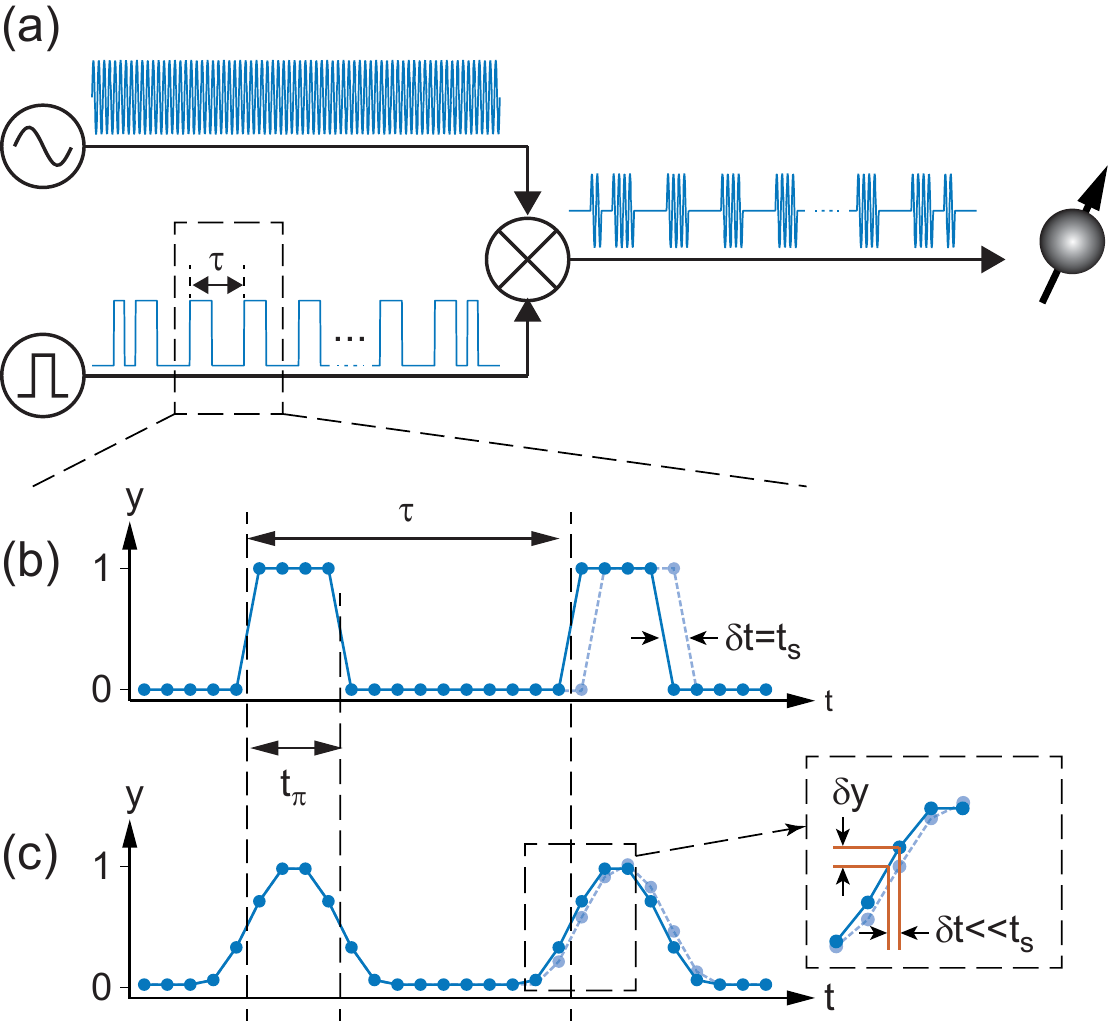}
\caption{\captionstyle
(a) Dynamical decoupling spectroscopy is based on a periodic pulse modulation of the qubit control field with a precisely timed pulse repetition time $\tau$.
(b) Modulation of the microwave signal with square pulses limits the time resolution to multiples of the sampling time $\ts$ of the pulse generator hardware.  Solid and faint profiles show original and time-shifted pulses.
(c) Shaped pulses, here with a cosine-square amplitude profile, enable much finer variations of the pulse timing at the same hardware sampling rate.  The minimum interpolated $\dt$ is set by the slope of the pulse envelope and the vertical resolution of the pulse generator (inset).
$\tpi$ is the duration of the $\pi$ pulse defined by the full width at half maximum of the pulse envelope.  In our experiments, the qubit is the solid-state spin of a single nitrogen-vacancy center in diamond. % and the control field has a frequency of $\sim 2.2\unit{GHz}$.
\label{fig:fig1}
}
\end{figure}

In this Letter we investigate the application of pulse shaping to enhance the timing resolution in the emerging field of dynamical decoupling spectroscopy \cite{cywinski08,alvarez11}.  Dynamical decoupling is a quantum control method developed to protect a quantum system against dephasing by environmental noise \cite{suter16}.  More recently dynamical decoupling sequences have also been applied to map out noise spectra and to detect time-varying signals with high signal-to-noise ratio \cite{alvarez11,kotler11,delange11,bylander11}.  In their simplest implementation, dynamical decoupling sequences consist of a periodic series of $\pi$ pulses with repetition time $\tau$ \cite{carr54} (see Figure \ref{fig:fig1}(a)).  For large numbers of pulses $N$, the spectral response of these sequences resembles that of a narrowband frequency filter, with center frequency $1/(2\tau)$ and bandwidth $1/(N\tau)$, that rejects noise at all frequencies except for those commensurate with the repetition time $\tau$.  By tuning $\tau$ into resonance with a signal at a particular frequency $\fac\approx 1/(2\tau)$, the sensitivity to the signal can be enhanced while suppressing the influence of noise, thereby resembling the properties of a classical lock-in amplifier in the quantum regime \cite{kotler11}.

When using many control pulses, the filter bandwidth becomes very narrow and the repetition time $\tau$ must be precisely tuned to the signal frequency $\fac \approx 1/(2\tau)$.  On any controller hardware, however, $\tau$ can only be adjusted in increments of the sampling time $\ts$.  This limits in practice the frequency resolution of the technique.  Specifically, when detecting a signal with frequency $\fac$, the minimum frequency increment is given by
\begin{equation}
\df = \frac{1}{2\tau} - \frac{1}{2\tau+2\ts} \approx 2\ts\fac^2 \ .
\end{equation}
Arbitrary waveform generators (AWGs) employed for the control of superconducting and spin qubits have sampling rates of typically $1\unit{GS/s}$, corresponding to a time resolution of $\ts=1\unit{ns}$.  When operating at high frequencies $\fac$ this timing resolution quickly becomes prohibitive.  For example, at a signal frequency of $\fac = 5\unit{MHz}$, the minimum frequency increment is $\df = 50\unit{kHz}$, which precludes the detection of weak signals with sharply defined spectra.  Although AWGs with faster sampling rates exist, they are extremely costly and barely reach adequate timing resolution.  Hardware sampling therefore is a severe limitation for dynamical decoupling spectroscopy.

Recently, an elaborate experimental scheme termed \textit{quantum interpolation} has been devised and demonstrated to overcome this issue \cite{ajoy17}. It enables a frequency sampling beyond the hardware limit of the control electronics by varying the interpulse delay between subblocks of the sensing sequence. This leads to an interpolation of the spin evolution at a more precisely controlled interpulse delay.

Here, we study the complementary and conceptually simpler approach of utilizing shaped control pulses to interpolate the pulse timing.  The concept of our method is illustrated in Figure \ref{fig:fig1}.  Instead of modulating the high frequency control field by the common square pulse profile (Fig. \ref{fig:fig1}(b)), we shape the envelope of the pulses by a smooth function.  In this study we use a cosine-square profile (Fig. \ref{fig:fig1}(c)), although any other smooth profile could be applied \cite{ernst90,steffen03}.  The pulse shape can be computed numerically before uploading the waveform data to the AWG and is therefore exceedingly simple to implement.  (Numerical code is given in the supplemental material \cite{supplemental}).  Because the AWG has a high vertical resolution, we can interpolate the center position of a pulse with a timing resolution $\dt$ that is far better than sampling time $\ts$.  The interpolated timing resolution $\dt$ is approximately given by the slope of the pulse envelope,
$\dt = (\partial y/\partial t)^{-1} \dy \sim \tpi \dy$,  where $\tpi$ is the duration of the pulse and $\dy$ the minimum amplitude increment.  Specifically, for a cosine-square shaped pulse of duration $\tpi=25\unit{ns}$ implemented on an AWG with 14 bits of vertical resolution ($\dy = 2^{-14}$), an interpolated timing resolution of $\dt \sim 1\unit{ps}$ can be expected.

To experimentally demonstrate the shaped-pulse interpolation method we study the coherent response of the electronic spin associated with a single nitrogen-vacancy (NV) center in a diamond single crystal.  Due to their excellent coherence properties, even at room temperature, NV centers have pioneered the area of applied quantum sensing, with applications in nanoscale magnetometry of condensed matter systems \cite{rondin14}, imaging of current distributions in nanostructures \cite{chang17} or structural magnetic resonance imaging of single proteins \cite{kong15,ajoy15}.  In particular for structural NMR imaging, very high frequency resolutions combined with MHz detection frequencies are required, providing a demanding test case for dynamical decoupling spectroscopy.
In our experiments, control pulses are generated on an AWG operating at 500 MS/s with 14 bits of vertical resolution (Tektronix AWG5002C), and upconverted to the $\sim 2.2\unit{GHz}$ qubit resonance by frequency mixing with a local oscillator (Fig. \ref{fig:fig1}(a)).  Additional amplification is used to reach Rabi frequencies of $\sim 20\unit{MHz}$ corresponding to pulse durations of $\tpi\sim 25\unit{ns}$.  Microwave pulses are applied to the NV center by a coplanar waveguide structure, and the NV spin state is initialized and read out by optical means \cite{loretz13}.  All experiments are performed under ambient conditions.
\begin{figure}[t]
\includegraphics[width=1\columnwidth]{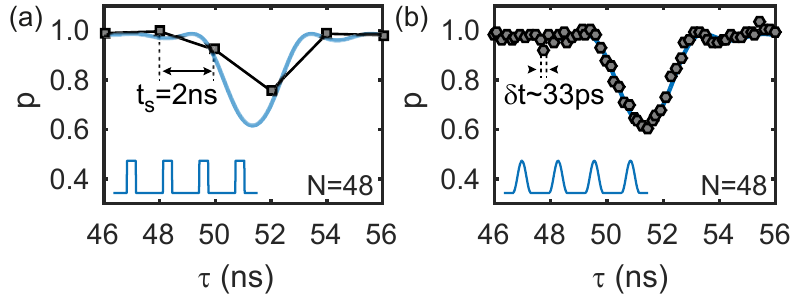}
\caption{\captionstyle
Experimental sampling resolution for a sensing sequence with (a) square control pulses and (b) cosine-square control pulses.  $\ts$ is the hardware sampling time and $\dt$ the interpolated sampling time enabled by the pulse shaping.  $N$ is the number of control pulses.  $p$ is the probability that the qubit sensor maintains its coherence for different values of the pulse repetition time $\tau$.  Solid lines show the theoretical response given by Eq. (2) and (3) with $\Bac$ as the only free parameter, squares and hexagons show the experimental data, and sketches show pulse shapes.
\label{fig:fig2}
}
\end{figure}

Figure \ref{fig:fig2} shows a first set of measurements, in which we directly compare the timing resolution of dynamical decoupling sequences with and without shaped pulses. For this purpose we combine the control field with a sinusoidal AC test signal ($\fac=9.746969\unit{MHz}, \Bac = 7.15\unit{\uT}$) before delivering it to the coplanar waveguide.  In this case the sensing sequence becomes resonant with the AC field for a pulse repetition time of $\tau=51.298\unit{ns}$.  When utilizing square pulses, the repetition time can only be stepped in increments of $\ts = 2\unit{ns}$ and the sampling of the AC signal spectrum is very coarse (Fig. \ref{fig:fig2}(a)).  In stark contrast, we can finely sample the spectrum using shaped control pulses and clearly augment the hardware-limited time resolution (Fig. \ref{fig:fig2}(b)).

\begin{figure}[b]
\includegraphics[width=1\columnwidth]{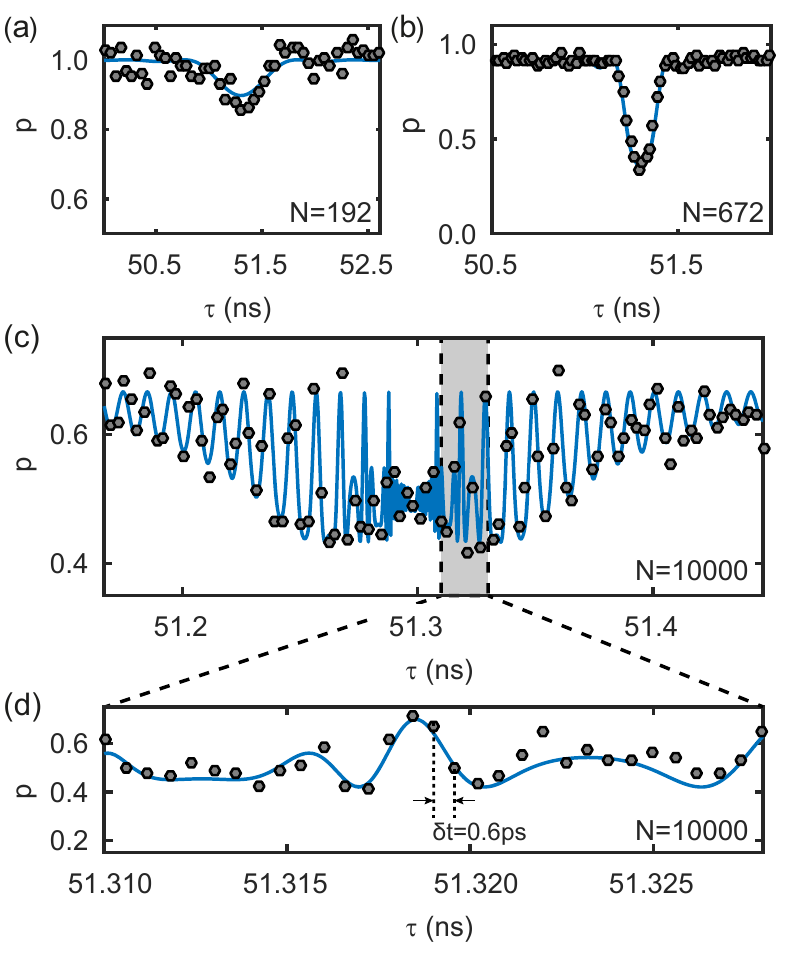}
\caption{\captionstyle
Sensor response to a dynamical decoupling sequence with (a) $N=192$, (b) $N=672$ and (c) $N=10,000$ shaped control pulses.  The frequency and amplitude of the AC test signal is $\fac=9.746969\unit{MHz}$ and $\Bac=0.84\unit{\uT}$ for all measurements, respectively.  (d) High-resolution plot of the $N=10,000$ dynamical decoupling sequence showing a timing resolution of $\dt=0.6\unit{ps}$. Solid lines reflect the theoretical model multiplied by an overall decoherence factor $\exp(-(t/T_2)^2)$ with $T_2 = 535\unit{us}$.
\label{fig:fig3}
}
\end{figure}
We have compared the experimental data to the expected spectral response for the dynamical decoupling sequence.  Because the phase of the AC magnetic field is not synchronized to the measurement sequence, we can describe the probability that the sensor qubit maintains its original state by \cite{degen16}
\begin{align}
p(t,\tau) &= \frac{1}{2} [1+ J_0 (W_{N,\tau} \gamma \Bac t)] \ .
\label{eq:transitionprob}
\end{align}
% To check: Brms (estimated only roughly based on settings of the frequency generator Vpp)
%
Here, $\gamma = 2\pi\times 28\unit{GHz/T}$ is the gyromagnetic ratio of the electronic sensor spin, $t=N \tau$ is the total duration of the phase acquisition, and $J_0$ is the zeroth-order Bessel function of the first kind.  Further, $W_{N,\tau}(\fac)$ is the spectral weighting or filter function of the sequence \cite{degen16},
\begin{align}
W_{N,\tau}(f_{\mathrm{ac}})
  &= \left| \frac{\sin(\pi f_{\mathrm{ac}} N \tau)}{\pi f_{\mathrm{ac}} N \tau} [1-\sec(\pi f_{\mathrm{ac}} \tau)] \right| \ ,
\label{eq:filterfunction}
\end{align}
which has a maximum response of $W=2/\pi$ when $\fac=1/(2\tau)$.  We find excellent agreement between the experimental spectra and the analytical filter profile of the dynamical decoupling sequence, but only the interpolated sequence provides the required fine frequency sampling.
We have verified that the filter profile does not depend on whether square or shaped control pulses are used \cite{supplemental}.

In a second set of experiments, shown in Figure \ref{fig:fig3}, we analyze the sensor response under the action of an increasing number of shaped microwave pulses. Here, we keep the frequency of the test signal unchanged but reduce its amplitude. We first record the response to a sequence with $N=192$ and 672 shaped pulses to calibrate the amplitude of the AC signal.  In both cases the sensor is operated in the small signal regime where the argument of the Bessel function is small ($W_{N,\tau} \gamma \Bac t \lesssim 2\gamma\Bac t/\pi \ll \pi$).

Subsequently, we tune the sensor into the strongly non-linear regime by increasing $N$ up to $10,000$ (Fig. \ref{fig:fig3}(c)).  For this large number of pulses, the argument of the Bessel function in Eq. \ref{eq:transitionprob} is no longer a small quantity because the total duration of the sequence $t=N\tau$ becomes very long.  The non-linear regime has recently been explored with trapped ions \cite{kotler13} and it has been found that this regime gives rise to spectral features far below the Fourier limit.  Here, we exploit these features to precisely characterize and test the frequency response of the sensor.  Figure \ref{fig:fig3}(d) shows a zoom into the center region of spectrum (c) that is acquired with a time increment of $\dt = 0.6\unit{ps}$.  We observe that even for this fine time resolution the observed response of the sensor spin agrees well with the model expressed by Eq. \ref{eq:transitionprob}.  The time increment of $0.6\unit{ps}$ corresponds to a frequency sampling of $114\unit{Hz}$, which is an improvement by over three orders of magnitude compared to the frequency sampling of $380\unit{kHz}$ possible without pulse shaping.

\begin{figure}[t]
\includegraphics[width=1\columnwidth]{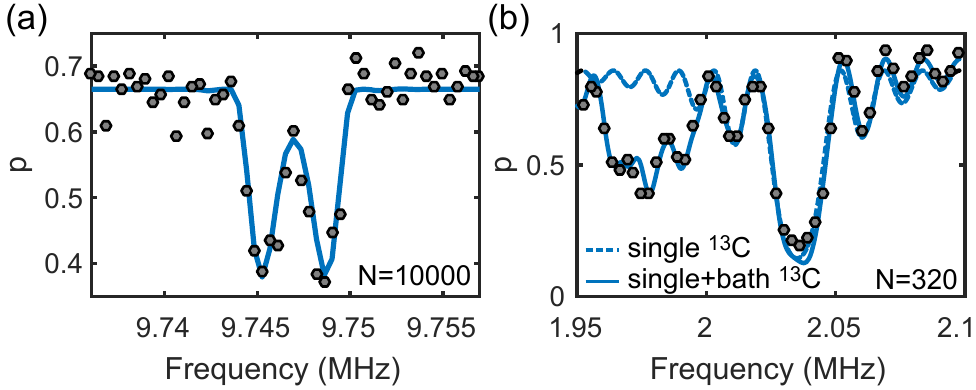}
\caption{\captionstyle
Applications of the method to high-resolution spectroscopy.
(a) Spectrum of two AC test signals separated by $3\unit{kHz}$.  Both signals can be clearly distinguished.
(b) NMR spectrum of \C nuclei located in close proximity to the NV center sensor spin.
Dashed line is the theoretical response to a single \C nucleus with hyperfine coupling parameters $\apar = 2\pi \times 114(1) \unit{kHz}$ and $\aperp = 2\pi \times 62(1) \unit{kHz}$.  Solid line includes three additional, more weakly coupled \C bath spins.
Frequency is $1/(2\tau)$.
\label{fig:fig4}
}
\end{figure}

To demonstrate the ability of the interpolated dynamical decoupling sequence to spectrally resolve nearby signals, we expose the sensor to a two-tone AC magnetic field composed of two slightly different frequencies.  We select equal amplitudes for both frequency components and operated the sensor in the linear regime.  As shown in Figure \ref{fig:fig4}(a) both frequency components can be clearly distinguished in the resulting spectrum even though the frequencies are only $3\unit{kHz}$ apart.

Finally, we demonstrate that the application of shaped pulses also enables the detection of NMR spectra with high frequency resolution.  Specifically, we detect the NMR signal of \C nuclei located in close proximity to the NV center \cite{taminiau12,zhao12,kolkowitz12}.  Figure \ref{fig:fig4}(b) shows the observed spectrum (points) for a sequence with $N=320$ control pulses together with a theoretical model (lines).  Because the response of the sensor spin is no longer described by the classical description of Eq. (\ref{eq:transitionprob}), we perform density matrix simulations of the NV-\C system to calculate the $p(t,\tau)$ response curve.  The simulations require knowledge of the parallel and perpendicular hyperfine coupling constants $\apar$ and $\aperp$, respectively, which we determine in separate high-resolution correlation spectroscopy measurements \cite{boss16}.  Two simulations are shown with Figure \ref{fig:fig4}(b): A first curve (dashed line) plots the expected response from the single \C nuclear spin.  The second curve (solid line) reflects a simulation that includes three additional, more weakly coupled \C nuclei.  The example of Figure \ref{fig:fig4}(b) clearly shows the advantage of a high sampling resolution for detecting NMR spectra.

To conclude, we have presented a simple method for greatly increasing the timing resolution of dynamical decoupling sequences.  Using sequences with up to 10,000 coherent, amplitude-shaped control pulses, we were able to improve the effective timing resolution of the interpulse delay by more than 3 orders of magnitude, with time increments as small as $0.6\unit{ps}$.  The resulting high frequency resolution has been demonstrated by sensing AC magnetic fields and NMR signals from individual carbon nuclear spins.
The method provides a simple technique to further push the boundaries in frequency resolution and sensitivity in quantum sensing applications and can also be applied to other physical implementations, such as other solid state defect spins, trapped ultracold atoms and ions, or superconducting qubits.
%The method will be useful for other applications beyond dynamical decoupling, such as for implementing fast phase gates with high fidelity.
%It can also be applied to other physical implementations, such as other solid state defects, trapped ultracold atoms and ions, or superconducting qubits.
%The method requires a minimum of numerical computation and can be directly implemented on existing controller hardware with an arbitrary waveform capability.

%%%%%%%%%%%%%% Acknowledgments

\begin{acknowledgments}
The authors thank Carsten Robens and Tobias Rosskopf for discussions and experimental support.  
This work was supported by Swiss NSF Project Grant $200021\_137520$, the NCCR QSIT, and the DIADEMS programme 611143 of the European Commission.
T.F.S. acknowledges Society in Science, The Branco Weiss Fellowship, administered by the ETH Zurich.
The work at Keio has been supported by KAKENHI (S) No.26220602 and JSPS Core-to-Core Program.
\end{acknowledgments}

%\bibliography{library}
%\bibliography{C:/Christian/ETH/labview/library/library}

\end{document}

\large
\begin{center}

\textbf{{Supplementary Information for: \\ ``High resolution quantum sensing with shaped control pulses''}}

\normalsize

\vspace{5 mm}

J. Zopes$^1$, K. Sasaki$^2$, K. S. Cujia$^1$, J. M. Boss$^1$, K. Chang$^1$, T. F. Segawa$^1$, K. M. Itoh$^2$, and C. L. Degen$^1$

\textit{$^1$Department of Physics, ETH Zurich, Otto Stern Weg 1, 8093 Zurich, Switzerland.}\\
\textit{$^2$School of Fundamental Science and Technology, Keio University, Yokohama 223-8522, Japan.}

\end{center}

\small

%%%%%%%%%%%%%%%%%%%%%%%%%%%%%%%%%%%%%%%%%%%%%%%%%%%%%%%%%

\subsection*{Computation of the cosine square pulse waveform}

The following Matlab code was used to calculate the amplitude modulation for a dynamical decoupling sequence with $\texttt{N}$ shaped pulses, as shown in Figure 1(a) and 1(c) of the main manuscript:
\begin{verbatim}
x = (0:n-1)*N/n;
x = mod(x,1);
x = 1-tau/tpi*abs(x-0.5);
x = max(x,0);
x = sin(pi*x/2).^2;
\end{verbatim}
Here, $\texttt{n}$ is the number of waveform samples, $\texttt{tau}$ is the interpulse delay, and $\texttt{tpi}$ is the duration of a $\pi$ pulse defined as the full width at half maximum of the envelope function.  Note that with this definition of $\texttt{tpi}$, the duration of a cosine-shaped $\pi$ pulse is equal to the duration of an equivalent square pulse with the same peak amplitude.  In our experiments, $\texttt{N}$ was up to 26,000, $\texttt{n}$ was up to about 500,000, and $\texttt{tpi}$ was about $25\unit{ns}$.

The resulting waveform envelope is then software multiplied with a carrier sine wave at $100\unit{MHz}$, and upconverted to the $\sim 2.2\unit{GHz}$ spin resonance by analog IQ mixing with the output of an external synthesizer.

\subsection*{Comparison of filter functions for dynamical decoupling sequences with different pulse shapes}

To investigate whether the type of control pulse used has an influence on the filter function of a dynamical decoupling sequence, we simulated the response of the sequence to a single \C nucleus using the density matrix method \cite{loretz15,boss16}.  For this example, we used the same parameters as for the \C in Fig. 4(b) of the main manuscript.  
The $\pi$ pulse duration was $\tpi = 25\unit{ns}$, the \C Zeeman frequency was $1.975\unit{MHz}$, and the detuning of the NV electronic spin due to the \NN nuclear was taken into account.

\begin{figure}[t]
\includegraphics[width=0.8\textwidth]{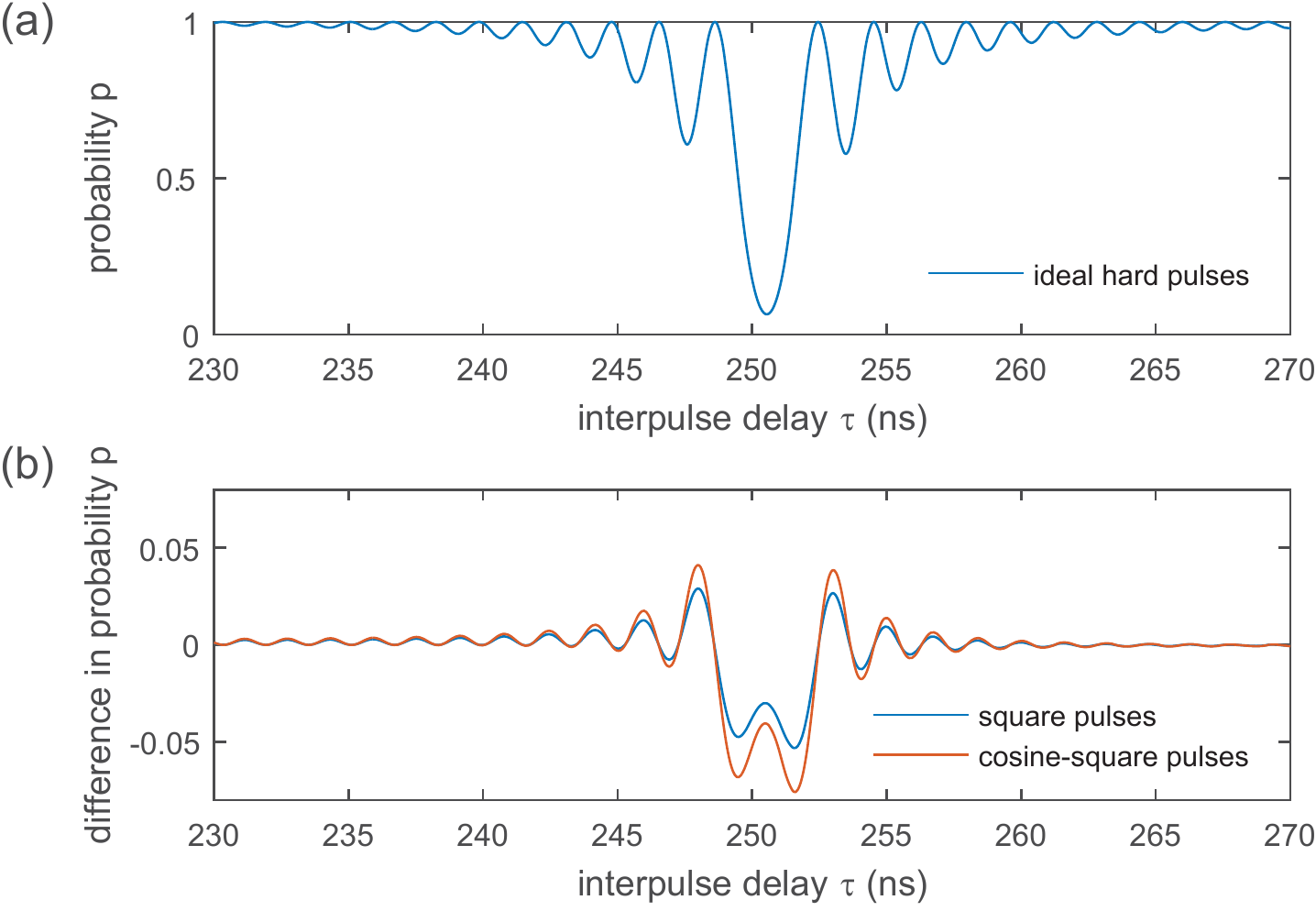}
\caption{Simulated response of a dynamical decoupling sequence with $N=320$ pulses to a single \C nuclear spin.
(a) Response of a sequence with ideal $\pi$ pulses. 
(b) Response of the sequences with square and cosine-square-shaped $\pi$ pulses, respectively.  Shown is the difference to the ideal response.
\label{fig:figS1}}
\end{figure}

We simulated the responses of sequences using four different control pulses: ideal, square, cosine-square-shaped, and cosine-square-shaped rounded to 14 bits.
Figure \ref{fig:figS1}(a) shows the response for the sequence with ideal (infinitely short and exact) $\pi$ rotations.  Figure \ref{fig:figS1}(b) plots the difference to the ideal response when using square pulses (blue curve) and when using cosine-square-shaped pulses (red curve).  We find that there is a difference to the ideal response, but that the difference between the square and cosine-square control pulses is small.  The differences are mainly due to the rather low Rabi frequency ($\sim 20\unit{MHz}$) which results in comparably long pulses and different phase pickup between ideal, square and cosine-square sequences.  We have also simulated the response to cosine-square pulses with a discrete amplitude, reflecting the 14 bits of vertical resolution of the AWG; here, the difference in $p$ is $<10^{-6}$ (not shown).
%

%%%%%%%%%%%%%%%%%%%%%%%%%%%%%%%%%%%%%%%%%%%%%%%%%%%%%%%%%%%%%%%%%%%%%%%%%%%%%%%%%%%%%%%%%%%%%%

%\bibliography{library}
%\bibliography{C:/Christian/ETH/labview/library/library}